# On-Chip Chemical Sensing Using Double-slot Silicon Waveguide

Sushma Gali, Akshay Keloth, and Shankar Kumar Selvaraja, *Senior Member, IEEE*

*Abstract*—In this paper, we present refractive index measurement using a double-slot silicon waveguide-based Mach Zehnder interferometer. We present a double-slot waveguide that offers the best sensitivity and limit of detection compared to wire and single-slot waveguides. We demonstrate ultra-low loss coupling between a single-mode waveguide and a double-slot waveguide and experimental proof for double-slot excitation. The double-slot waveguide is used to demonstrate a highly sensitive concentration sensor. An unbalanced Mach-Zehnder is used as the sensor device to demonstrate concentrations of Potassium Chloride in deionised water. A sensitivity of 700 nm/RIU and a limit of detection (LOD) of $7.142 * 10{-6}$ RIU is achieved experimentally. To the best of our knowledge, the demonstrated sensitivity is the highest for an on-chip guided waveguide sensing scheme.

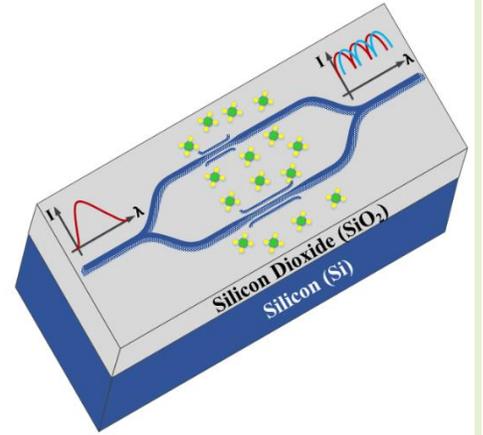

## I. INTRODUCTION

INTEGRATED optical sensors offer an excellent platform for sensing in medical diagnostics, food quality checks, environment monitoring, etc., and provide the possibility for incorporating lab on a chip. These lab on-chips are usually made to provide multiple optical functionalities [1]-[5]. For that, they need to be compact and sensitive sensors and are usually manufactured on high-index-contrast platforms such as silicon-on-insulator (SOI). Additionally, these chips must be disposable to avoid complex cleaning processes and should be available for lower prices. SOI platform is majorly used for integrated photonic applications as it is compatible with the matured complementary metal oxide semiconductor (CMOS) technology that allows mass fabrication and provides cost-effective sensors [3]. In integrated optical sensors, sensing is accomplished by the interaction of the evanescent field of the propagating mode with the cladding [4],[6]. Refractive index variation in the top cladding environment can be evaluated using a resonant/interferometric device by monitoring the spectral response. Primary photonic devices like ring resonators and Mach Zehnder Interferometers (MZI) [9]-[13],[4] can be used to track shifts in the spectral response. However, strip-based ring resonators and MZIs offer minimal spatial field overlap with the top cladding [6]. Hence, the sensitivity provided by strip waveguide-based devices is low, i.e., in the order of a few tens of nanometers. On the other hand, slot waveguides offer high light-matter interaction and, thus, high sensitivity [7]. Unlike strip waveguides, where most of the light is confined in a high refractive index medium, in slot waveguides, light is in the low refractive index medium and thus offers high light-matter interaction[14],[15]. The evanescent field available for sensing in slot waveguides is 25% greater than a strip waveguide. Besides the evanescent field, sensitivity is also related to the group index, as shown in equation (1),

$$Sensitivity\ S = \frac{\Delta\lambda}{\Delta n_{clad}} = \frac{\lambda_r}{n_g}\left(\frac{dn_{eff}}{dn_{clad}}\right)...(1)$$

Where $\lambda_r$ is the resonant wavelength, $n_g$ is the group index, $dn_{eff}$ is the change in the effective refractive index of the slot waveguide with the change in top cladding and $dn_{clad}$ is the change in the refractive index of the top cladding. The group index of a double-slot waveguide configuration is relatively lower than a single-slot waveguide and hence can provide high sensitivity. In this paper, we proposed an efficient and compact coupling scheme to couple light to a double-slot waveguide from a strip waveguide and exploited its sensing ability using an asymmetric MZI structure. Waveguide configuration is optimized to achieve maximum sensitivity using finite difference simulation [8]. Double-slot mode excitation is initially verified using the group index extracted from the spectral response of MZI. Later, the device is used as a refractive index sensor by drop-casting aqueous solutions of potassium chloride (KCl). The estimated sensitivity is experimentally verified with the spectral shift observed from an asymmetric MZI. We confirmed the spectral shift due to refractive index change with an Abbe refractometer. We report the highest sensitivity for a waveguide-based on-chip sensor and also estimate micro-molar sensing capability.

## II. DOUBLE-SLOT WAVEGUIDE OPTIMIZATION

The dimensions of the double-slot waveguide are optimized to achieve maximum sensitivity. As shown in equation (1), sensitivity depends on the resonant wavelength ($\lambda_r$), group index ($n_g$), $dn_{eff}/dn_{clad}$ and intuitively, evanescent field fraction (EFF). The EFF is the fraction of the total electric field of the propagating wave that interacts with the top cladding in the sensing region. EFF can be formulated as

$$\text{EFF} = \frac{n_g \iint_A \varepsilon |E|^2 \, dx \, dy}{n_{clad} \iint_{tot} \varepsilon |E|^2 \, dx \, dy} \cdots (2)$$

Where $|E|^2$ is the energy density of the electric field, $\varepsilon$ is the relative permittivity, and $n_{clad}$ is the refractive index of the top cladding. A is the area where the electric field interacts with top cladding over the total (tot) area where the electric field is extended across the transverse cross-section of the waveguide. Figure 1 (a) and (b) show the electric field distribution of the fundamental TE mode in a strip and double-slot waveguide. The refractive indices considered for the simulation are 3.45 for silicon and 1.44 for the buried silicon dioxide (bottom cladding). As it appears, almost 80% of light is confined in the strip waveguide core, and thus a minimal amount of electric field is available for sensing. The maximum sensitivity that can be extracted using a strip waveguide is 70 nm/RIU [4]. Unlike strip waveguides, slot waveguides can confine 40-50% of the propagating light in the low- refractive index cladding and are a promising candidate for bulk refractive index sensing [15]. By extending from a single slot to a double, the interaction with the cladding can offer higher sensitivity [21]. The sensitivity of the slot waveguide would be maximum, if the slope $dn_{eff}/dn_{clad}$ is high, the group index $n_g$ is low, and the maximum possible electric field in the top cladding. Figure 2(a) shows the effect of slot waveguide rail width for a fixed gap of 90 nm on the EFF. The group index ($n_g$) of the double-slot waveguide with a propagating TE slot mode extracted from the FDM simulation is 1.76. The slope ($dn_{eff}/dn_{clad}$), i.e., a change in the waveguide effective refractive index with a change in the top cladding (sensing environment) is extracted using FDTD simulation and is shown in figure 2(b). A sensitivity of 705 nm/RIU can be achieved with the optimized slot waveguide dimensions of rail width ($w_s$ = 200nm) and gap (g = 90nm).

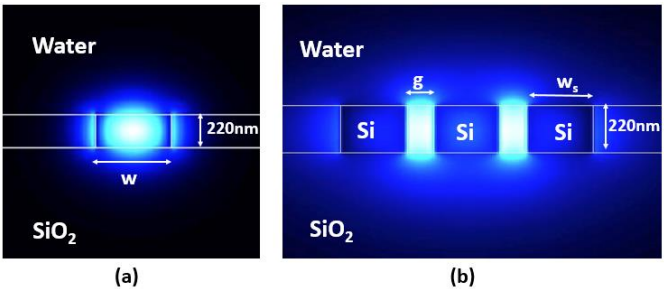

Fig. 1. Electric field distribution of fundamental TE mode extracted from the 2D FDM simulations with water as the top cladding; (a) Strip waveguide

(width (W) = 400 nm, thickness = 220 nm). (b) Double slot waveguide of dimensions;(rail width ($W_s$) = 200 nm and gap (g) = 90 nm, thickness = 220 nm).

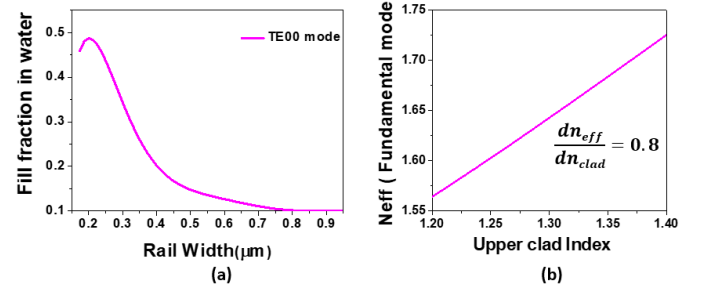

Fig. 2. (a) EFF of double slot waveguide at various slot rail widths ($W_s$) (b) Change in the effective refractive index of the double slot waveguide due to change in the top cladding.

## III. STRIP TO DOUBLE SLOT COUPLING

### A. Coupler Design

Though the double-slot waveguide offers better light-matter interaction for sensing, it comes with the challenges of coupling and propagation loss [17]-[19]. The sidewall roughness created during the fabrication process is the main reason for the high-propagation loss in a slot waveguide. The loss could be reduced using high-resolution patterning technology [16]. However, the major challenge is coupling light efficiently to a double-slot waveguide. In this section, we demonstrate a compact and efficient strip to a double-slot waveguide coupler and validate slot mode excitation. The proposed strip-to-double-slot coupling scheme is shown in figure 3. A double-slot waveguide is formed by a tapered central waveguide, and two curved waveguides are used to adiabatically couple light from a silicon strip waveguide to a double-slot waveguide. The taper and arcs together provide an adiabatic transition of the fundamental strip waveguide mode to the fundamental double slot mode. Figure 4 (a) presents the FDTD electrical field simulation for the proposed scheme. The smooth transition of the E-field in the direction of propagation is shown in series figures 2(b) to 2(e). The dependency of coupling efficiency on the length of the taper structure and arc radius is shown in figure 4(f).

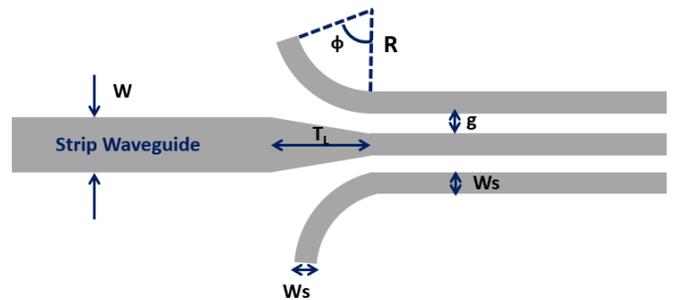

Fig. 3. Simulated Proposed structure to couple light to a double slot waveguide; Strip waveguide (width (W) = 400 nm). Double slot waveguide dimensions; (width of slot rail ($W_s$) = 200 nm, gap (g) = 90 nm, taper length($T_L$) = 4 µm, arc radius (R) = 10 µm and angle (φ) = 45°

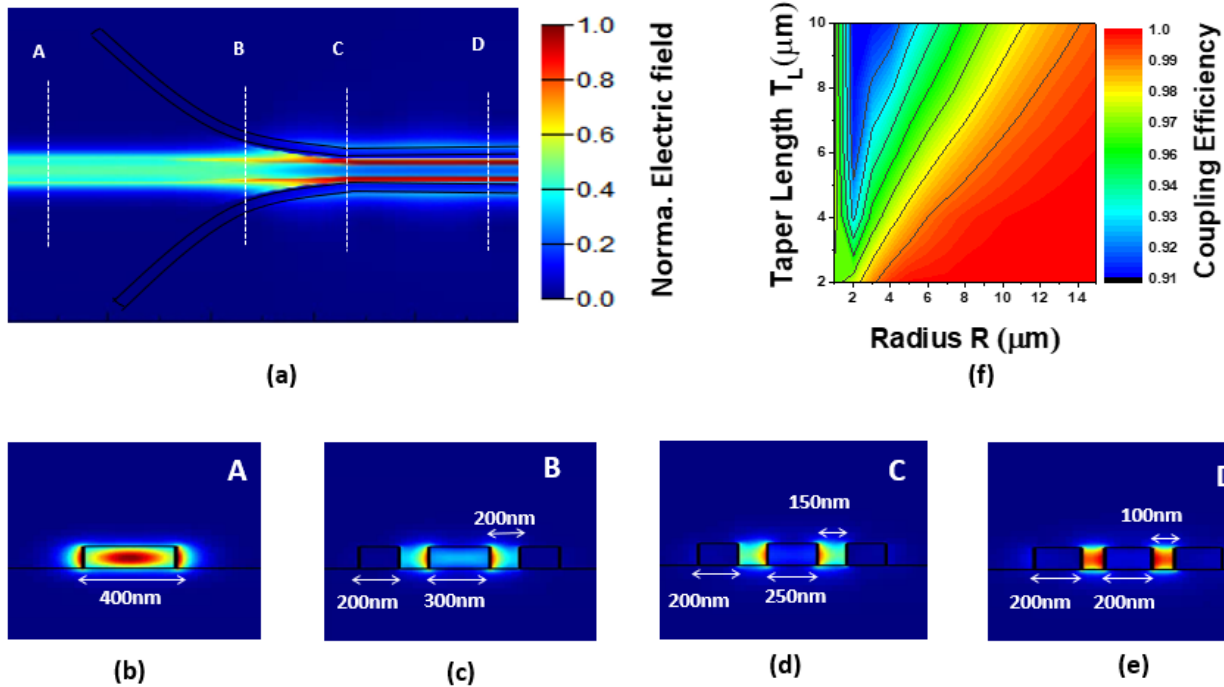

A slot rail width of 200 nm and gap of 90 nm is the optimized dimensions of a single mode double-slot waveguide to couple light from a 450 nm wide single-mode silicon strip waveguide. For the above-mentioned dimensions of the double slot waveguide, the optimized coupler design will have a taper length of 4 µm and an arc radius of 10 µm. We kept the angle φ shown in Figure 3 at 45 degrees to design the lithography patterning file easily. One can also vary φ along with taper length and arc radius to get a compact design.

### B. Fabrication

A standard SOI wafer with a 220 nm silicon device layer on a 2 µm buried oxide is used to fabricate the double-slot-based waveguide. The test structure is made of a grating in-out coupler, 450 nm access wire waveguide, strip-to-double-slot and double-slot-to-strip coupler, and double-slot waveguide of various lengths. The pattering of the waveguides and gratings are done using electron-beam lithography and a dry etch process. Waveguides are etched 220 nm deep, while the gratings are defined by a 70 nm shallow etch. Figure 5 shows the optical and scanning electron microscope (SEM) image of a fabricated device. The excitation and propagation of the double-slot mode are validated by using a Mach-Zhender interferometer (MZI). The MZI has double slot waveguides in both arms with an arm imbalance length of 200 µm (ΔL). The spectral response of the MZI depends on the group index of the propagating mode in the MZI waveguide. By using the spectral transmission characteristics of the MZI, one could validate slot-mode excitation and propagation. Figure 6 shows the layout of the MZI test device and SEM images of a fabricated device.

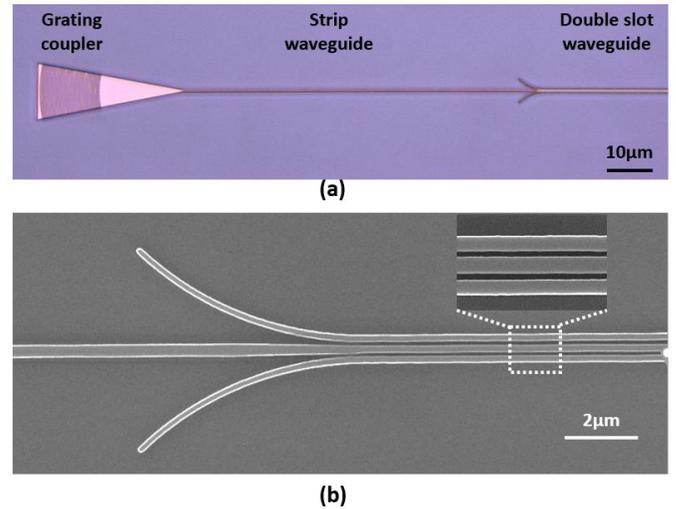

### C. Optical Characterization and Double Slot Mode Validation

The optical transmission characteristics of the fabricated devices were measured using a 1550 nm broadband SLED with a 3-dB bandwidth of 50 nm and an optical spectrum analyzer. Figure 7 shows the spectral response obtained from a double-slot waveguide-based asymmetric MZI. The group index ($n_g$) of the double slot waveguide is calculated from the free spectral range

$(FSR = \frac{\lambda^2}{n_g \Delta L})$ of the MZI spectral response. The experimental group index (1.73) and simulated group index (1.76) are in good agreement and thus confirm slot mode excitation.

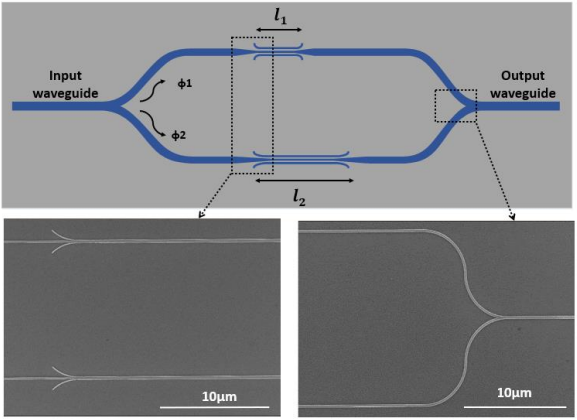

Fig. 6. Double slot-based asymmetric MZI structure

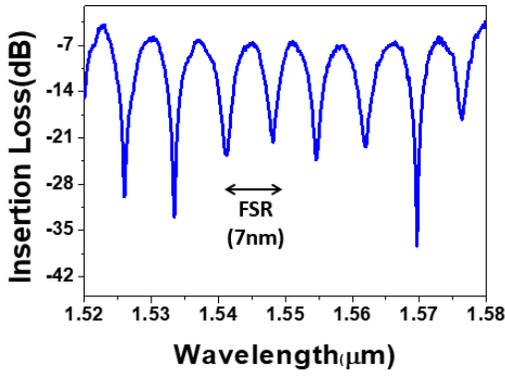

Fig.7. The spectral response obtained from a double slot-based asymmetric MZI (ΔL=200 μm)

### D. Propagation loss and Coupler efficiency

The propagation loss of the double-slot waveguide is estimated using the cut-back method. A propagation loss of 6.17±0.6 dB/mm is measured for a double-slot waveguide rail width of 200 nm and a gap of 90 nm. Figure 8 shows the propagation loss measurement. The loss is primarily attributed to sidewall scattering, as there are six sidewalls. For sensing applications, the loss can be compensated by the interaction volume. Though the loss is high, the desired functionality could be achieved.

The absolute transmission through the slot waveguide depends on efficient coupling between the wire and the slot. An insertion loss of 0.025 dB/coupler is measured by normalizing strip and double slot waveguide loss. The ultra-low coupling loss shows near-ideal power transfer between the wire and the double-slot waveguide. The efficient excitation of double slot mode with low insertion loss makes it a promising structure to exploit fluid sensing.

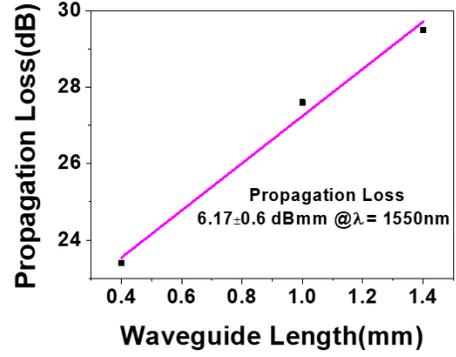

Fig.8. Propagation loss of double slot waveguide estimated using cut-back method.

## IV. FLUID SENSING USING DOUBLE SLOT WAVEGUIDE

In this section, we present fluid sensing using a double-slot waveguide-based asymmetric MZI. As explained in section 3(c), initially, slot mode excitation is validated using an asymmetric MZI with double-slot waveguides. For sensing demonstration, the imbalance (ΔL) is increased 400 μm. We extracted an experimental group index (1.7) for the double-slot waveguide from the spectral response of the asymmetric MZI, and it is close to the simulated group index of 1.75. To explore fluid sensing, we drop casted potassium chloride (KCl) solution in three different concentrations 50M,100M, and 150M.

For the chosen MZI configuration (figure 7), we can write the phase of the propagating mode as follows

$$\phi 1 = \frac{2\pi}{\lambda}(n_s l_1 + n_w(l_2 - l_1))$$

$$\phi 2 = \frac{2\pi}{\lambda}(n_s l_2)$$

$$\Delta\phi = \frac{2\pi}{\lambda}(n_s - n_w)\Delta L$$

Where ϕ1 is the phase in the upper arm of the MZI, ϕ2 is the phase in the lower arm of the MZI, $n_s$ is the effective refractive index of slot region, $n_w$ is the effective refractive index of the strip region, and $\Delta L$ is the slot length difference. For a given $\Delta L$, which is always positive, the effective index difference between double-slot and strip waveguide $(n_s - n_w)$ is negative. We observed that $n_w$ is always greater than $n_s$ for a chosen strip and double-slot waveguide dimensions. Hence, $|n_s - n_w|$ reduces with increasing $n_{clad}$. The refractive index of the top cladding $n_{clad}$ increases with an increase in fluid concentration. Thus, we observed a blue shift with increasing concentration of KCl, which is observed experimentally in Figure 9(a). Sensitivity $(\Delta\lambda/\Delta n_{clad})$ is calculated using the spectral shift of the asymmetric MZI and the change in the refractive index of the KCl solutions. The slope $\frac{dn_{eff}}{dn_{clad}} = 0.8$ shown in figure 2(b) is extracted from the simulation. We achieved a maximum sensitivity of 700 nm/RIU, which is close to our theoretical estimation presented in Section 2. We also calculated the Limit of detection (LOD) [22] as the ratio of spectral resolution to spectral sensitivity.

$$Limit\ of\ Detection\ LOD = \frac{\delta\lambda}{S}$$

Where $\delta\lambda$ is the wavelength resolution of the detector and $S$ is the spectral sensitivity from equation (1). We experimentally achieved a LOD of $2.85\text{X}10^{-5}$ RIU. We will also attribute that if we use a wavelength sampling resolution of 5 pm, a LOD of $7.142\text{X}10^{-6}$ RIU can be achieved, enabling on-chip micromolar concentration detection. The extracted refractive index from the spectral measurement is verified with the Abbe Refractometer data. The refractometer measures the refractive index of the drop-casted solution by the principle of total internal reflection [25]. Figure 9(b) shows a linear fit for the refractive index measurements with the Abbe refractometer and on-chip. The experimentally measured sensitivity is compared with the state-of-the-art on-chip fluid sensing on the SOI platform, which is presented in Table I. Although photonic Crystal structures can offer high sensitivity compared to slot waveguide-based devices, they come with complicated fabrication and implementation [24],[25].

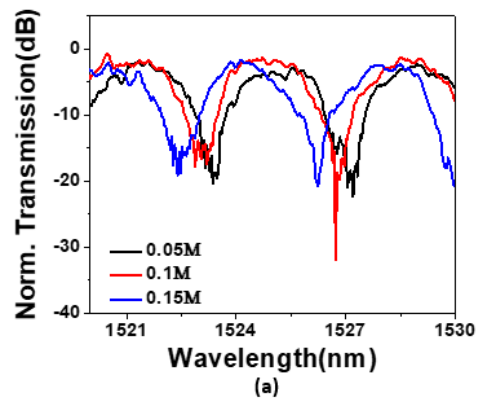

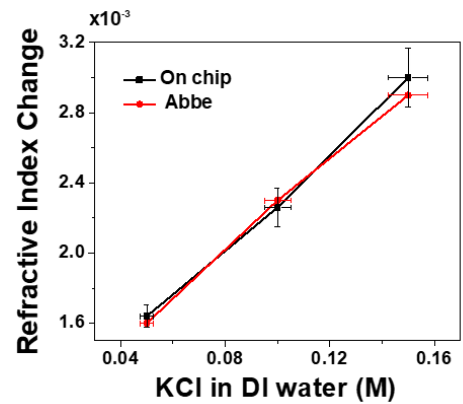

Fig.9. (a) Spectral response of the fabricated MZI with drop cast KCl (b) Refractive Index measurement Abbe Vs. On-chip

## TABLE I
### Slot-Based Refractive Index Sensing Platform for Fluid Sensing

| Device | Waveguide | System resolution ($\delta\lambda$ (pm)) | Sensitivity (nm/RIU) | System LOD (RIU) | Ref |
|---|---|---|---|---|---|
| MZI | Strip | NA | 460 ($2\pi$) | $3.0 \times 10^{-5}$ | [4] |
| Ring | Strip | 5 | 70 | $7.1 \times 10^{-5}$ | [26] |
| | Slot | 12.5 | 298 | $4.2 \times 10^{-5}$ | [27] |
| | | 12.0 | 345 | $3.4 \times 10^{-5}$ | [28] |
| | Slot | 5.0 | 476 | $1.1 \times 10^{-5}$ | [20] |
| PhC | Slot | NA | 1538 | $7.0 \times 10^{-6}$ | [23] |
| | | NA | 510 | $1.0 \times 10^{-5}$ | [24] |
| **MZI** | **Double Slot** | **5.0** | **700** | **$7.1\text{x}10^{-6}$** | **This work** |

## V. CONCLUSION

In conclusion, we presented a detailed study of the strip-to-double-slot waveguide coupling scheme and fluid sensing using double-slot waveguides. We experimentally demonstrated a compact and efficient strategy to couple light between a strip waveguide and a double-slot waveguide. The slot mode excitation is verified using spectral obtained from an asymmetric MZI. We efficiently coupled light to a double-slot waveguide with an ultra-low insertion loss of 0.025 dB/coupler and a propagation loss of 6.1±70.6 dB/mm. We also demonstrated refractive index sensing using a silicon double-slot waveguide-based Mach-Zehnder interferometer. An asymmetric MZI is used to track the refractive index change. We experimentally measured a sensitivity of 700nm/RIU with a limit of detection (LOD) of $7.142\text{X}10^{-6}$ RIU. To the best of our knowledge, we demonstrate the best in class sensitivity and LOD for an index-guided waveguide scheme.

## ACKNOWLEDGMENT

We acknowledge funding support from MoE, MeitY, and DST through NNetRA. We also thank the UGC for PhD fellowship and Prof. Ramakrishna Rao chair fellowship.

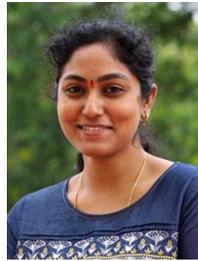
**Sushma Gali** received the B.Tech. Degree in Electronics and Communication from Yogi Vemana University, India, in 2012, and the M.Tech. Degree in VLSI from Vaagdevi Institute of Technology, India, in 2014. She is currently pursuing a Ph.D. degree in Photonics at the Centre for Nano Science and Engineering, Indian Institute of Science, Bengaluru, India. Her current research interests include Photonic integrated sensing, On-chip Spectroscopy, and process integration.

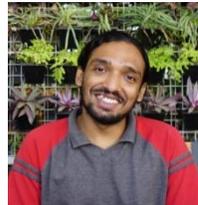
**Akshay Keloth** completed his Integrated Master's in Photonics from Cochin University of Science and Technology in 2018. Between 2019-2022 he worked as a Project Associate at the Centre for Nano Science and Engineering (CeNSE), Indian Institute of Science, Bangalore. He si currently a graduate researcher at the University of Twente, The Netherlands. His current research interests include Silicon and Silicon nitride Photonic IC for sensing and nonlinear applications

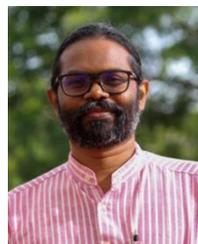
**Shankar Kumar Selvaraja** Graduated with a PhD degree degree in Photonics Engineering from Ghent University—IMEC, Ghent, Belgium, in 2011. He received MEngg and MS degree from Anna University, India and the University of Twente, The Netherlands. He was a Postdoctoral Researcher with IMEC from 2011 to 2013 and a Process Integration Engineer of Silicon Photonics with IMEC from 2013 to 2014. Since March 2014, he has been with the Centre for Nanoscience and Engineering, Indian Institute of Science, Bengaluru, India where he is currently an Associae Professor. His research group works in the area of photonics integrated circit enabled communication, computing and sensing.